\pdfoutput=1
\RequirePackage{ifpdf}
\ifpdf 
\documentclass[pdftex]{sigma}
\else
\documentclass{sigma}
\fi

\numberwithin{equation}{section}

\newtheorem{Theorem}{Theorem}[section]
\newtheorem{Lemma}[Theorem]{Lemma}

\usepackage{mathtools}
\usepackage[braket, qm]{qcircuit}

\newcommand{\ea}[1]{\begin{align}#1\end{align}}

\newcommand{\nn}{\nonumber \\}
\newcommand{\la}{\langle}
\newcommand{\ra}{\rangle}
\newcommand{\mmod}{\mathrm{~mod~}}

\newcommand{\mbb}{\mathbb}

\newcommand{\bmp}{\begin{pmatrix}}
\newcommand{\emp}{\end{pmatrix}}
\newcommand{\bmb}{\begin{bmatrix}}
\newcommand{\emb}{\end{bmatrix}}

\DeclareMathOperator{\tr}{Tr}

\begin{document}
\allowdisplaybreaks

\newcommand{\arXivNumber}{2312.13400}

\renewcommand{\PaperNumber}{044}

\FirstPageHeading

\ShortArticleName{SICs and the Triangle Group $(3,3,3)$}

\ArticleName{SICs and the Triangle Group $\boldsymbol{(3,3,3)}$}

\Author{Danylo YAKYMENKO}

\AuthorNameForHeading{D.~Yakymenko}

\Address{Institute of Mathematics of NAS of Ukraine, Kyiv, Ukraine}
\Email{\href{mailto:danylo.yakymenko@gmail.com}{danylo.yakymenko@gmail.com}}
\URLaddress{\url{https://imath.kiev.ua/~dandan/}}

\ArticleDates{Received January 02, 2024, in final form May 10, 2024; Published online May 29, 2024}

\Abstract{The problem of existence of symmetric informationally-complete positive op\-er\-a\-tor-valued measures (SICs for short) in every dimension is known as Zauner's conjecture and remains open to this day. Most of the known SIC examples are constructed as an orbit of the Weyl--Heisenberg group action. It appears that in these cases SICs are invariant under the so-called canonical order-three unitaries, which define automorphisms of the Weyl--Heisenberg group. In this note, we show that those order-three unitaries appear in projective unitary representations of the triangle group $(3,3,3)$. We give a full description of such representations and show how it can be used to obtain results about the structure of canonical order-three unitaries. In particular, we present an alternative way of proving the fact that any canonical order-three unitary is conjugate to Zauner's unitary if the dimension $d>3$ is prime.}


\Classification{20C35; 81P15; 81R05}

\section{Introduction}

Let $H = \mbb{C}^d$ be a complex Hilbert space of finite dimension $d$.
A \textit{positive operator-valued measure $($POVM for short$)$} is
a measure whose values are positive semi-definite Hermitian matrices
$M_i$ on $H$. That is,
$M_i^\dag = M_i \geqslant 0$ and
$\sum_{i=1}^n M_i = I$
if the set $\{M_i\}$ has a finite size $n$.

A POVM is a model of a general measurement that one can perform on a quantum system
with $d$ degrees of freedom. Matrices $M_i$ correspond to distinguishable outcomes
labelled by the indices $i$. If $\rho$ is a density matrix on $H$,
which encodes a state of the quantum system, then the outcome labelled $i$
appears with probability
\[
{\rm Prob}(i) = \tr(\rho M_i)
\]
in a measurement event.
The von Neumann (projective or PVM) measurements are a special case
of POVMs, where each $M_i$ is the projection onto the $i$-th subspace in
an orthogonal decomposition $H = \oplus_i H_i$.

A \textit{symmetric informationally-complete POVM $($SIC-POVM or just SIC$)$}
is a POVM $\{M_i\}$ of size $n=d^2$, where each matrix $M_i=\frac{1}{d}\op{v_i}{v_i}$
has rank one and for each $i\neq j$, we have
$
\operatorname{Tr}(M_i M_j) = {\rm const} = \frac{1}{d+1}.
$
It corresponds to a set of $d^2$ pure states $\ket{v_i} \in H$ such that
\begin{gather*}
|\ip{v_i}{v_j}|^2 = \frac{1}{d+1}, \quad i\neq j,
\qquad
\sum_{i=1}^{d^2} \op{v_i}{v_i} = d I.
\end{gather*}

By knowing the probabilities $\operatorname{Tr}(\rho M_i) = \la \rho, M_i\ra_{\rm HS}$ of the outcomes $\{i\}$
(here ${\rm HS}$ stands for the Hilbert--Schmidt inner product),
we can completely reconstruct the quantum state $\rho$, which justifies the naming.
Note that it is not possible with a POVM of a lesser size $n < d^2$.
In a real scenario of measuring copies of $\rho$ multiple times, we get $\operatorname{Tr}(\rho M_i)$
as an estimate based on the frequency of outcome $i$.
This gives us an estimate $\hat{\rho}$ of $\rho$.
One can show that the expectation of~${||\rho - \hat{\rho}||^2_{\rm HS}}$,
averaged over all $\rho$ which are unitary equivalent to a fixed one,
is minimized exactly for SICs amongst other minimal informationally-complete POVMs \cite{Scott}.
This makes SICs the best choice for
a~``standard'' unbiased measurement in quantum mechanics.

The problem is that we do not know how to construct them in general.
Initially, SICs' existence in every dimension $d$ was conjectured
by Zauner in \cite{Zauner} (and independently by Renes et al.\ in~\cite{Renes}).
Since then a vast body of research was carried out
(e.g., \cite{Aligned2d, Appleby, Aligned, n^2+3, n^2+3=4p, Yard, BosWal, Fla, Fuchs, Kopp, Scott2, SG, Zhu}).
Either exact or numerical solutions have been
found in all dimensions $d \le 193$ and some others. The current state-of-the-art
was achieved in \cite{n^2+3, n^2+3=4p}, where authors proposed a recipe for finding SICs
in dimensions of the form $d=n^2+3$ and verified it in a number of dimensions,
including~${d=39604}$. However, it remains unproven in general,
and it is also related to unsolved problems in number theory.

Most known SICs are constructed as an orbit of the action
of the \textit{Weyl--Heisenberg group} ${\rm WH}(d)$.
It is a group generated by the \textit{clock and shift matrices} defined by
\[
Z\ket{k} = \omega^k \ket{k},
\qquad X\ket{k} = \ket{k+1},
\]
respectively, where $\omega = \exp \bigl(\frac{2\pi {\rm i}}{d}\bigr)$
and $\{\ket{k} \mid  k \in \mbb{Z}_d\}$ is the standard basis in $H$.

For a unit vector $\ket{f} \in H$, the set $\bigl\{ X^kZ^l\ket{f} \mid k,l \in \mbb{Z}_d\bigr\}$ is called a
\textit{WH SIC} if it is a SIC, and $\ket{f}$ is called a \textit{fiducial vector} in such a case.

In any dimension $d$, there exists an order-three unitary $\mathfrak{Z}$, called \textit{Zauner's unitary},
which gives an automorphism of ${\rm WH}(d)$ under the conjugation (see Section~\ref{sec:wh} for details).
The strong version of Zauner's conjecture states that in every dimension $d$ there exists
a fiducial vector $\ket{f}$ which is an eigenvector of $\mathfrak{Z}$.
This means that the corresponding WH SIC is invariant under the action of $\mathfrak{Z}$.
Moreover, the collected evidence suggests that any WH SIC is invariant under a
\textit{canonical order-three unitary}, which is related to $\mathfrak{Z}$ anyway.
This makes it essential for SICs' constructions, yet we do not have an explanation for this either.
Canonical order-three unitaries are also important in other problems, see, e.g., \cite{BBCH}.

In this paper, we show that $\mathfrak{Z}$ along with the group ${\rm WH}(d)$
appear in projective unitary representations of the ordinary triangle group $\Delta_{(3,3,3)}$,
which can be abstractly defined by generators and relations
\[
\Delta_{(3,3,3)} = \big\la a,b,c \mid a^3=b^3=c^3=1,\,  abc=1\big\ra.
\]
Such representations were first obtained in \cite{RL},
but here we give their simplified description with a modified proof in Section~\ref{sec:tri}.
We then show how to use it to investigate properties of canonical order-three unitaries
in Section~\ref{sec:can}. In particular, we give an alternative way of proving the fact that any
canonical order-three unitary is conjugate to Zauner's unitary if the dimension $d>3$ is prime.
We hope that the symmetric nature of
$\Delta_{(3,3,3)}$ could provide insights into the importance of $\mathfrak{Z}$ for SIC constructions.

\section{Weyl--Heisenberg group and its symmetries}\label{sec:wh}

In this section, we mainly follow Appleby \cite{Appleby}.
Recall that the Weyl--Heisenberg group is generated by
the clock and shift matrices $Z$, $X$:
\[
Z = \bmp
1 & 0 & 0 & \cdots & 0\\
0 & \omega & 0 & \cdots & 0\\
0 & 0 & \omega^2 & \cdots & 0\\
\vdots & \vdots & \vdots & \ddots & \vdots\\
0 & 0 & 0 & \cdots & \omega^{d-1}
\emp,
\qquad
X = \bmp
0 & 0 & \cdots & 0 & 1\\
1 & 0 & \cdots & 0 & 0\\
0 & 1 & \cdots & 0 & 0\\
\vdots & \vdots & \ddots &\vdots &\vdots\\
0 & 0 & \cdots & 1 & 0\\
\emp,
\]
where $\omega = \exp \bigl(\frac{2\pi{\rm i}}{d}\bigr)$.
They satisfy $Z^d = X^d = I$ and a kind of commutation relation $ZX = \omega XZ$.
It is easy to see that each element of the generated group has the
form $\omega^mX^kZ^l$, where $k,l,m \in \mbb{Z}_d$.
For further constructions, it is convenient to also add the factor $\tau = -\exp \bigl(\frac{\pi {\rm i}}{d}\bigr)$ to~it.
We thus define the group by ${\rm WH}(d) = \la X, Z, \tau I \ra$.
Note that $\tau^2 = \omega$ and if $d$ is odd then \smash{$\tau = \omega^{\frac{d+1}{2}}$}
so it adds nothing, but doubles the group in even dimensions.
In any case, for the quotient over the group center, denoted by $\la \xi I \ra$, we have ${\rm WH}(d) / \la \xi I \ra \simeq \mbb{Z}_d \times \mbb{Z}_d$.

The $d^2$ \textit{displacement matrices} for ${\bf a} = (a_1,a_2) \in \mbb{Z}_d^2$ defined by
\[
D_{\bf a} = \tau^{a_1a_2} X^{a_1}Z^{a_2} \in {\rm WH}(d)
\]
are cosets representatives of the quotient group ~${{\rm WH}(d) / \la \xi I \ra}$.
They satisfy the relations $D_{\bf a} D_{\bf b} = \tau^{\la \bf a, b\ra} D_{\bf a+b}$,
where $\la {\bf a, b} \ra = a_2b_1 - b_2a_1$ is the symplectic form.
In particular, $D_{\bf a}^\dag = D_{-\bf a}$.

The Clifford group ${\rm Cli}(d)$ is defined as the group of unitary automorphisms
of the Weyl--Heisenberg group. That is,
\[
{\rm Cli}(d) = \bigl\{ A \in {\rm Uni}(d) \mid A^\dag {\rm WH}(d) A = {\rm WH}(d) \bigr\}.
\]
Its importance comes from the fact that if $\ket{f}$ is a WH SIC fiducial vector,
then so is $A\ket{f}$ (for a~different SIC, in general).
Thus WH SIC solutions are naturally grouped in orbits of the Clifford group action.

The ${\rm Cli}(d)$ group has the following structure.
Let $\bar{d} = d$ if $d$ is odd and $\bar{d} = 2d$ otherwise.
By~${\rm SL}(2, \mbb{Z}_{\bar{d}} )$, we denote the group of $2 \times 2$ symplectic matrices
$M = \left(\begin{smallmatrix} m_1 & m_2 \\ m_3 & m_4 \end{smallmatrix}\right)$ with
entries from~$\mbb{Z}_{\bar{d}}$ and condition ${\rm det}(M) = 1$.
For a symplectic matrix $M$,
its \textit{symplectic unitary} is defined (up to a scalar factor) by
\[
A_M = \sum_{r,s \in \mbb{Z}_d}\tau^{m_2^{-1}(m_1 s^2 -2rs + m_4 r^2)} \op{r}{s}
\]
if $m_2^{-1}$ (a multiplicative inverse) exists in $\mbb{Z}_{\bar{d}}$. Otherwise,
one has to decompose $M$ into a product of matrices in which $m_2^{-1}$ exists
and then take the product of its symplectic unitaries.

The group ${\rm SL}(2, \mbb{Z}_{\bar{d}} )$ acts naturally on $\mbb{Z}_d^2$ as
on vectors of size $2$, therefore we can define
the semidirect product ${\rm SL}(2, \mbb{Z}_{\bar{d}}) \ltimes \mbb{Z}_d^2$ with respect to this action.
It can be checked that $\forall M \in {\rm SL}(2, \mbb{Z}_{\bar{d}})$, $\forall {\bf a} \in \mbb{Z}_d^2$
\[
A_M D_{\bf a} A_M^\dag = D_{M{\bf a}}.
\]
This implies $A_M \in {\rm Cli}(d)$.
Define $A_{[M, {\bf b}]} = D_{{\bf b}} A_M$, ${\bf b} \in \mbb{Z}_d^2$.
It was proved \cite[Theorem 1]{Appleby} that the map
$[M, {\bf b}] \longrightarrow A_{[M, {\bf b}]}$
is a surjective homomorphism from
${\rm SL}(2, \mbb{Z}_{\bar{d}}) \ltimes \mbb{Z}_d^2$ to ${\rm Cli}(d) / \la \xi I \ra$,
it is an isomorphism in odd dimensions and has a kernel of size $8$ in even dimensions.

Note that $A_{[M, {\bf 0}]} = A_M$ just permutes the set of displacement matrices,
while $A_{[M, {\bf a}]}$ with ${\bf a} \neq {\bf 0}$ also adds phases to them.
For $A \in {\rm Cli}(d)/ \la \xi I \ra$, by
$M_A \in {\rm SL}(2, \mbb{Z}_{\bar{d}})$ we denote
a symplectic matrix such that $A = A_{[M_A, {\bf a}]}$
for some ${\bf a} \in \mbb{Z}_d^2$ ($M_A$ is not unique in even dimensions,
but this will be irrelevant to us).

A prominent example of an automorphism of ${\rm WH}(d)$ is the discrete Fourier transform matrix
\[
F = \frac{1}{\sqrt{d}}\sum_{r,s \in \mbb{Z}_d}\omega^{rs} \op{r}{s} = \frac{1}{\sqrt{d}}\bmp
1 & 1 & 1 & \cdots & 1\\
1 & \omega & \omega^2 & \cdots & \omega^{d-1}\\
1 & \omega^2 & \omega^4 & \cdots & \omega^{2(d-1)}\\
\vdots & \vdots & \vdots & \ddots & \vdots\\
1 & \omega^{d-1} & \omega^{2(d-1)} & \cdots & \omega^{(d-1)(d-1)}\\
\emp.
\]
The corresponding symplectic matrix of $F$ is
$M_F = \left(\begin{smallmatrix} 0 & -1 \\ 1 & 0 \end{smallmatrix}\right)$, $F = A_{M_F}$.
It exchanges the clock and shift matrices,
$FXF^\dag = Z$, $FZF^\dag = X^{-1}$, and has order four for $d>2$.

On the other hand, Zauner's unitary $\mathfrak{Z}$ has the symplectic matrix
$M_\mathfrak{Z} = \left(\begin{smallmatrix} 0 & -1 \\ 1 & -1 \end{smallmatrix}\right)$, $\mathfrak{Z} = A_{M_\mathfrak{Z}}$,
and up to a scalar equals to
\[
\mathfrak{Z} = \frac{{\rm e}^{\frac{{\rm i}\pi (d-1)}{12}}}{\sqrt{d}} \sum_{r,s=0}^{d-1} \tau^{r^2+2rs}\ket{r}\bra{s}
= {\rm e}^{\frac{{\rm i}\pi (d-1)}{12}}\bmp
1 & 0 & 0 & \cdots & 0\\
0 & \tau & 0 & \cdots & 0\\
0 & 0 & \tau^4 & \cdots & 0\\
\vdots & \vdots & \vdots & \ddots & \vdots\\
0 & 0 & 0 & \cdots & \tau^{(d-1)^2}\\
\emp F,
\]
where the multiplier ${\rm e}^{\frac{{\rm i}\pi (d-1)}{12}}$ ensures that $\mathfrak{Z}^3 = I$.
The matrix $\mathfrak{Z}$ permutes ${\rm WH}(d)$ by the rules
\[
\mathfrak{Z}X\mathfrak{Z}^\dag = Z,
\qquad
\mathfrak{Z}Z\mathfrak{Z}^\dag = \tau X^{-1}Z^{-1},
\qquad
\mathfrak{Z}D_{(a_1, a_2)}\mathfrak{Z}^\dag = D_{(-a_2, a_1-a_2)}.
\]

For a unitary $A \in {\rm Cli}(d)$, its \textit{Clifford trace} is defined by
$\tr_{{\rm c}}(A) =\tr(M_A) \bmod d$.
This is well-defined because it does not depend on the choice of $M_A$.
Clearly, Zauner's unitary and its conjugates in the Clifford group have Clifford trace $-1$.
In general, if $\tr_{{\rm c}}(A)=-1$ then $A$ is of order three
(i.e., we can pick a~phase such that $A^3=I$
and we always assume so in what follows),
unless $A=I$ in dimension three. Such unitaries are called \textit{canonical order-three unitaries}.

While Zauner conjectured that a WH SIC fiducial vector $\ket{f}$ can be
found as an eigenvector of~$\mathfrak{Z}$, the collected evidence suggests
that in any case $\ket{f}$ is an eigenvector of a~canonical order-three unitary.
This observation became known as the Appleby--Zauner conjecture.
It was proved in \cite{Fla} that any canonical order-three unitary is conjugate
to Zauner's if the dimension $d>3$ is prime. It is not true in general.
Theorem~9.1 in \cite{BosWal}
gives the characterization of their conjugacy classes. They are given by the sets of
representatives
\begin{gather*}
 \{\mathfrak{Z}\}, \quad d\neq0 \mmod 3, \qquad
 \bigl\{\mathfrak{Z}, \mathfrak{Z}^2\bigr\}, \quad d=0 \mmod 9 \quad\text{or} \quad d=3, \\
 \bigl\{\mathfrak{Z}, \mathfrak{Z}^2, A_{M_1}\bigr\}, \qquad d=3\mmod 9 \quad\text{and}\quad d \neq 3, \\
 \bigl\{\mathfrak{Z}, \mathfrak{Z}^2, A_{M_2}\bigr\}, \qquad d=6\mmod 9, \\
\end{gather*}
where
\[
M_1 = \bmp 1 & 3 \\ \frac{4d-3}{3} & -2 \emp,
\qquad
M_2 = \bmp 1 & 3 \\ \frac{2d-3}{3} & -2 \emp \mod \bar{d}.
\]

Note that the conjugacy class $\{M_1\}$ corresponds to $F_a$ in \cite{Scott2,SG}, where
WH SIC fiducial vectors, which are eigenvectors of $A_{M_1}$, found for all small $d = 3(3k+1)$.
For some reason, no WH SIC fiducials are found as eigenvectors of $A_{M_2}$ when $d = 3(3k+2)$.

In addition to Clifford unitaries, we can consider anti-unitary operators that also
define ${\rm WH}(d)$ automorphisms. Let $\hat{J}$ be the anti-unitary operator that
acts on $H$ by the rule
\[
\hat{J}\bigg(\sum_{r \in \mbb{Z}_d} c_r \ket{r}\bigg) = \sum_{r \in \mbb{Z}_d} \bar{c}_r \ket{r}.
\]
Its conjugate action on ${\rm WH}(d)$ is described by
\[
\hat{J} D_{(a_1, a_2)} \hat{J}^\dag = D_{(a_1,-a_2)}, \qquad \forall (a_1,a_2) \in \mbb{Z}_d^2.
\]

In fact, the disjoint union
\[
{\rm ECli}(d) = {\rm Cli}(d) \cup \bigl\{\hat{J}A \mid A \in {\rm Cli}(d)\bigr\}
\]
forms the \textit{extended Clifford group}.
In this group a product of two anti-unitaries is a unitary operator,
while anti-unitary times unitary remains anti-unitary.
If we denote by ${\rm ESL}(2, \mbb{Z}_{\bar{d}})$ the group of symplectic matrices
with ${\rm det}(M) = \pm 1$, then by \cite[Theorem 2]{Appleby}
the extended map~${[M, {\bf a}] \longrightarrow D_{\bf a}A_M}$ is a surjective
homomorphism from
${\rm ESL}(2, \mbb{Z}_{\bar{d}}) \ltimes \mbb{Z}_d^2$
to ${\rm ECli}(d) / \la \xi I \ra$,
under which~$\hat{J}$ corresponds to $\left(\begin{smallmatrix} 1 & \hphantom{-}0 \\ 0 & -1 \end{smallmatrix}\right)$
and anti-unitaries in general to $M$ with ${\rm det}(M) = -1$.

\section{Triangle group and its representations}\label{sec:tri}

\begin{figure}[h!]\centering
\includegraphics[height=60mm]{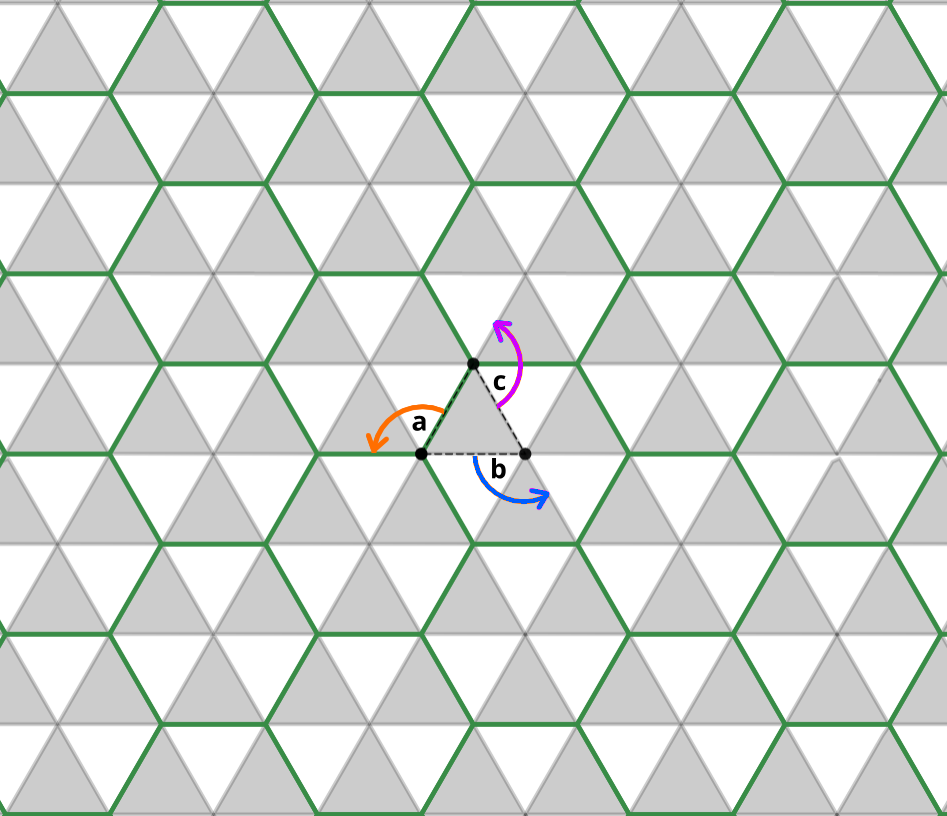}
 \caption{The realization of $\Delta_{(3,3,3)}$ as motions of the plane.} \label{fig:abc}
 \end{figure}

The \textit{ordinary triangle group} $\Delta_{(3,3,3)}$
(such groups are also known as von Dyck groups)
can be defined abstractly by the generators and relations
\[
\Delta_{(3,3,3)} = \big\la a,b,c \mid a^3=b^3=c^3=1,\, abc=1\big\ra.
\]
It has a nice realization as motions of the real plane \cite{Magnus}.
Elements $a$, $b$, $c$ can be represented by $2\pi/3$ rotations around the vertices
of an equilateral triangle with side length 1, see Figure~\ref{fig:abc}.

The elements $u=a^2b$, $v=aba$, $w=ba^2=u^{-1}v^{-1}$ correspond to the translations
of the pattern (see Figure~\ref{fig:abc}) by one step of length $\sqrt{3}$
in the directions north $\uparrow$,
southwest $\swarrow$ ($240^\circ$ north) and
southeast $\searrow$ ($120^\circ$ north), respectively.
They generate the abelian subgroup $\la u,v,w \ra \simeq \mbb{Z} \times \mbb{Z}$
of translations.
Note that the horizontal shift of length $1$ preserves the alternating triangles pattern,
but not the additional hexagonal. This is not an element of~$\Delta_{(3,3,3)}$.

One can see that
$\Delta_{(3,3,3)}/\la u,v,w \ra \simeq \mbb{Z}_3$.
Moreover,
\[
\Delta_{(3,3,3)} \simeq \mbb{Z}_3 \ltimes \la u,v,w \ra
\simeq \mbb{Z}_3 \ltimes (\mbb{Z} \times \mbb{Z}),
\]
where $\mbb{Z}_3$ acts by shifting the generators $u \rightarrow v \rightarrow w \rightarrow u$.
It follows from the identities $a^{-1}ua = v$, $a^{-1}va = w$ and $a^{-1}wa = u$.

We are interested in projective unitary representations of $\Delta_{(3,3,3)}$.
Denote the set of complex units by ${\rm S}^1 = \{\xi \in \mbb{C} \mid |\xi| = 1 \}$.

A \textit{projective unitary representation} of a group $G$
is a homomorphism from $G$ to the group
${\rm PUni}(n) = {\rm Uni}(n)/\la \xi I\ra$ of projective unitaries.
Alternatively, it is a map $\pi\colon G \longrightarrow {\rm Uni}(n)$ that satisfies
\begin{align*}
\pi(xy) = \pi(x)\pi(y) \cdot \xi(x, y),
\qquad
\xi(x, y) \in {\rm S}^1.
\end{align*}

Let $\pi$ be any such representation of $\Delta_{(3,3,3)}$.
It is easy to see that $(\pi(a))^3, (\pi(b))^3, (\pi(c))^3 \in \bigl\{ \gamma I \mid \gamma \in {\rm S}^1\bigr\}$.
Hence, we can choose scalars $\xi_a, \xi_b, \xi_c \in {\rm S}^1$ such that
\ea{\label{tri:ABC}
A^3 = B^3 = C^3 = I,\qquad ABC = \lambda I,
}
where $A = \xi_a\pi(a)$, $B = \xi_b\pi(b)$, $C = \xi_c\pi(c)$ and $\lambda \in {\rm S}^1$.
Henceforth, we focus on solving equation~\eqref{tri:ABC}.

Recall that two solutions $(A,B,C)$ and $(A',B',C')$
are unitary equivalent if and only if
\[
U(A,B,C)U^\dag = \bigl(UAU^\dag,UBU^\dag,UCU^\dag\bigr) = (A',B',C')
\]
for some $U \in {\rm Uni}(n)$.
A unitary representation $(A,B,C)$ is irreducible if and only if its not equivalent to
$(A',B',C') \oplus (A'',B'',C'') = (A'\oplus A'',B'\oplus B'',C'\oplus C'')$.
Equivalently, there is no proper subspace which is invariant for all $A$, $B$, $C$.

In fact, the irreducible unitary representations of equation~\eqref{tri:ABC} were
classified by Livinskyi and Radchenko in \cite{RL}. But here we present a modified and more
symmetric exposition of the results and also show the relation with Zauner's unitary.
Additionally, note that there is a~general (Mackey's) theory of finding representations
of virtually abelian groups~\cite{Kyr}.

By computing the determinants of expressions from equation~\eqref{tri:ABC}, one can conclude that $\lambda^{3n} = 1$,
so~$\lambda$ is a root of unity.
Denote $\mu = \lambda^3$ and let
\begin{align*}
U = A^2B,
\qquad
V = ABA,
\qquad
W = BA^2.
\end{align*}
From equation~\eqref{tri:ABC}, it follows that
\ea{ \label{tri:UVW}
UV = \mu VU,
\qquad
VW = \mu WV,
\qquad
WU = \mu UW,
\qquad
UVW = \mu I,
\qquad
WVU = I.
}

The key idea is that we find irreducible representations of the equation \eqref{tri:UVW}
at first.
Then we find how $(U,V,W)$, that comes from an irreducible solution $(A,B,C)$,
can split into irreducible parts.
The representation theory of the relation $UV=\mu VU$ is well known \cite{Davidson,Weyl}.
If $\mu = \exp (2\pi k{\rm i}/d)  = \omega^k$, ${\rm gcd}(k,d)=1$,
then there are only $d$-dimensional irreducible representations which can be given by
\[
U_\alpha = \alpha Z^k,
\qquad
V_\beta = \beta X,
\]
where $\alpha, \beta \in {\rm S}^1$ are parameters of the representation.
Two such representations are equivalent if and only if
$\bigl(\alpha'^d, \beta'^d\bigr) = \bigl(\alpha^d, \beta^d\bigr)$. Thus the pair $\bigl(\alpha^d, \beta^d\bigr)$
specifies a representation uniquely.
Packed together, the irreducible $d$-dimensional solutions for $U$, $V$, $W$ in equation \eqref{tri:UVW}
can be described by
\begin{gather}
  U_\alpha = \alpha Z^k = \alpha D_{(0,k)},\qquad V_\beta = \beta X = \beta D_{(1,0)}, \qquad
 W_\gamma = \gamma \tau^k X^{-1}Z^{-k} = \gamma D_{(-1,-k)},\label{tri:UVWsol}
\end{gather}
where $\alpha \beta \gamma = \tau^k$.
And thus $\alpha^d \beta^d \gamma^d = \tau^{dk} = (-1)^{k(d-1)}$.

Two representations $(U_{\alpha'}, V_{\beta'}, W_{\gamma'})$ and $(U_\alpha, V_\beta, W_\gamma)$
are equivalent if and only if
\[
\bigl(\alpha'^d, \beta'^d, \gamma'^d\bigr) =\allowbreak \bigl(\alpha^d, \beta^d, \gamma^d\bigr).
\]

Let us return to the initial problem of finding irreducible solutions $(A,B,C)$ of equation~\eqref{tri:ABC}.
In general, the triple $(U,V,W) = \bigl(A^2B, ABA, BA^2\bigr)$, up to equivalence,
decomposes into a direct sum of irreducible summands given by equation~\eqref{tri:UVWsol}.
It corresponds to a decomposition of $H$ into invariant subspaces of $(U,V,W)$,
although this decomposition may not be unique if the summands are equivalent.
Let $H_1 \subseteq H$ be one of the invariant subspaces of dimension $d$.
Then the subspaces $H_2=A(H_1)$ and $H_3=A^2(H_1)$ are also invariant for $(U,V,W)$,
because $U(H_2) = \bigl(AVA^{-1}\bigr)(H_2) = AV(H_1) = A(H_1) = H_2$ and similarly for $H_3$ and $V, W$.
Since $H_1 + H_2 + H_3$ is invariant for $A$ as well as for $U$, $V$, $W$ we immediately obtain
that irreducible $(A,B,C)$ are at most~$3d$-dimensional.

Let us be more specific. Pick an orthonormal basis in $H_1$ so that
\[
 (U,V,W)_{|_{H_1}} = (U_{\alpha},V_{\beta},W_{\gamma}),
\]
for some parameters $(\alpha, \beta, \gamma)$ as in equation~\eqref{tri:UVWsol}.
This also fixes bases in $A(H_1)$ and $A^2(H_1)$ accordingly.
Then
\begin{align*}
 & (U,V,W)_{|_{H_2}} = \bigl(A(V,W,U)A^{-1}\bigr)_{|_{A(H_1)}}
 = (V,W,U)_{|_{H_1}} = (V_{\beta},W_{\gamma},U_{\alpha}), \nn
 & (U,V,W)_{|_{H_3}} = \bigl(A^2(W,U,V)A^{-2}\bigr)_{|_{A^2(H_1)}}
 = (W,U,V)_{|_{H_1}} = (W_{\gamma},U_{\alpha}, V_{\beta}).
\end{align*}

Consider the symplectic matrix
\[
M_{\mathfrak{Z}_k} = \bmp 0 & -k^{-1} \\ k & -1 \emp \in {\rm SL}(2, \mbb{Z}_{\bar{d}} )
\]
along with the symplectic unitary $\mathfrak{Z}_k \in {\rm Cli}(d)$ it generates
(we take a scalar such that $\mathfrak{Z}_k^3=1$).
$\mathfrak{Z}_k$ has order three and permutes $X \rightarrow Z^k \rightarrow D_{(-1,-k)} \rightarrow X$.
Hence,
\begin{align*}
 & \mathfrak{Z}_k(V_{\beta},W_{\gamma},U_{\alpha})\mathfrak{Z}_k^\dag
 = (U_{\beta},V_{\gamma},W_{\alpha}),
 \qquad \mathfrak{Z}_k^2(W_{\gamma},U_{\alpha},V_{\beta})\mathfrak{Z}_k^{2\dag}
 = (U_{\gamma},V_{\alpha},W_{\beta}).
\end{align*}

What we get is that the shift
$(U_{\alpha}, V_{\beta},W_{\gamma}) \rightarrow (V_{\beta},W_{\gamma},U_{\alpha})$
in a solution \eqref{tri:UVWsol} corresponds to
the change of parameters
$({\alpha}, {\beta}, {\gamma}) \rightarrow ({\beta}, {\gamma}, {\alpha})$.
Three shifts $(U_{\alpha}, V_{\beta},W_{\gamma})$, $(V_{\beta},W_{\gamma},U_{\alpha})$,
$(W_{\gamma},U_{\alpha},V_{\beta})$ are inequivalent unless
$\alpha^d = \beta^d = \gamma^d$, which can happen only in three cases
where \smash{$\alpha^d = \sqrt[3]{(-1)^{k(d-1)}}$} for three cubic root choices.
If shifts are inequivalent, then~$H_1$,~$H_2$,~$H_3$ should be orthogonal to each other
according to the representation theory.
In this case, the representation of $(A,B,C)$ is unique for
$\bigl({\alpha^d}, {\beta^d}, {\gamma^d}\bigr)$ and can be written as
\ea{\label{tri:abcsol}
A = \bmp
0 & 0 & I \\
I & 0 & 0 \\
0 & I & 0 \\
\emp,
\qquad
B = \bmp
0 & 0 & W_\gamma \\
U_\alpha & 0 & 0 \\
0 & V_\beta & 0 \\
\emp,
\qquad
C = \lambda\bmp
0 & 0 & U^\dag_{\alpha} \\
V^\dag_{\beta} & 0 & 0 \\
0 & W^\dag_{\gamma} & 0 \\
\emp,
}
in the chosen above basis in $H_1 \cup A(H_1) \cup A^2(H_1)$, with
\[
 U = \bmp
 U_\alpha & 0 & 0 \\
 0 & V_\beta & 0 \\
 0 & 0 & W_\gamma \\
 \emp,
 \quad
 V = \bmp
 V_\beta & 0 & 0 \\
 0 & W_\gamma & 0 \\
 0 & 0 & U_\alpha \\
 \emp,
 \quad
 W = \bmp
 W_\gamma & 0 & 0 \\
 0 & U_\alpha & 0 \\
 0 & 0 & V_\beta \\
 \emp.
\]

Let us consider the remaining situation
where \smash{$\alpha^d = \beta^d = \gamma^d = \sqrt[3]{(-1)^{k(d-1)}}$},
which we call \textit{the singularity case}.
We can assume that \smash{$\alpha = \beta = \gamma = \sqrt[3]{\tau^k}
= \omega_3^l \exp\bigl(\frac{\pi {\rm i} k(d+1)}{3d}\bigr)$}, $l=0,1,2$,
where $\omega_3 = \exp\bigl(\frac{2\pi{\rm i}}{3}\bigr)$. Define
\begin{align*}
& H_1' = \bigl\{\ket{f} + A^{-1}\mathfrak{Z}_k\ket{f} + A^{-2}\mathfrak{Z}_k^{2}\ket{f} \mid  \ket{f} \in H_1 \bigr\} \subset H, \\
 &H_2' = \bigl\{\ket{f} + \omega_3A^{-1}\mathfrak{Z}_k\ket{f} + \omega_3^2A^{-2}\mathfrak{Z}_k^{2}\ket{f} \mid  \ket{f} \in H_1 \bigr\} \subset H, \\
& H_3' = \bigl\{\ket{f} + \omega_3^2A^{-1}\mathfrak{Z}_k\ket{f} + \omega_3A^{-2}\mathfrak{Z}_k^{2}\ket{f} \mid  \ket{f} \in H_1 \bigr\} \subset H.
\end{align*}
One can see that $H_1'$ is invariant for $A$
and
\begin{align*}
 U(\ket{f} + A^{-1}\mathfrak{Z}_k\ket{f} + A^{-2}\mathfrak{Z}_k^{2}\ket{f})
 &= U \ket{f} + A^{-1}W\mathfrak{Z}_k\ket{f} + A^{-2}V\mathfrak{Z}_k^{2}\ket{f} \\
 &= U_\alpha \ket{f} + A^{-1}W_\alpha\mathfrak{Z}_k\ket{f} + A^{-2}V_\alpha\mathfrak{Z}_k^{2}\ket{f}\\
& = U_\alpha \ket{f} + A^{-1}\mathfrak{Z}_kU_\alpha\ket{f} + A^{-2}\mathfrak{Z}_k^{2}U_\alpha\ket{f}
 \in H_1',
\end{align*}
so it is invariant for $U$ and thus for $V$ and $W$ too. The same holds
for $H_2'$, $H_3'$. Note that $H_1'+H_2'+H_3' = H_1+H_2+H_3$ and $\dim H_i'$ is
either $d$ or $0$ for each $i$ due to invariance.
Hence, $\dim H_i' = d$ for at least one index $i$.
But this means that irreducible representations of $(A,B,C)$
are at most $d$-dimensional in this case and we should have $H_1=H_2=H_3$
from the beginning.
It follows that~$A$ is an automorphism of $\la U, V, W\ra = \la U_\alpha, V_\alpha, W_\alpha\ra$
since $V = A^{-1}UA$, $W = A^{-2}UA^2$.
From the description of ${\rm Cli}(d)$, we know that $A$ coincides with $\mathfrak{Z}_k$ up to $\omega_3$ factor.
We conclude that there are three inequivalent representations
\begin{gather}
 A = \omega_3^j \mathfrak{Z}_k,
 \qquad B = \omega_3^j \mathfrak{Z}_k U_\alpha,
\qquad C = \lambda \omega_3^j U_\alpha^\dag \mathfrak{Z}_k,\nonumber \\
 U = U_\alpha,
\qquad V = V_\alpha,
\qquad W = W_\alpha,\label{tri:aaasol}
\end{gather}
where $j=0,1,2$ for each \smash{$\alpha = \sqrt[3]{\tau^k}$}, so that is 9 in total.
Moreover, the representation \eqref{tri:abcsol} decomposes exactly into
the three parts that correspond to the values of $j=0,1,2$ each, where the invariant subspaces are
$H'_{j+1}$.
We summarize this in the following theorem, which refines \cite[Theorem~5]{RL}:
\begin{Theorem}\label{tri:thm}
Let $\mu = \lambda^3 = \exp (2\pi k {\rm i}/d) = \omega^k$, ${\rm gcd}(k,d)=1$.
Consider the set of parameters~${(\alpha, \beta, \gamma) \in {\rm S}^1 \times {\rm S}^1 \times {\rm S}^1}$
with $\alpha\beta\gamma = \tau^k$ and
the equivalence relation $\bigl(\alpha'^d, \beta'^d, \gamma'^d\bigr) = \bigl(\alpha^d, \beta^d, \gamma^d\bigr)$.
Then equations \eqref{tri:abcsol}
define inequivalent irreducible $3d$-dimensional representations of \eqref{tri:ABC},
except three cases where \smash{$\alpha^d = \beta^d = \gamma^d = \sqrt[3]{(-1)^{k(d-1)}}
= \omega_3^l (-1)^{k(d-1)}$}, $l=0,1,2$.
In each of those cases, equations~\eqref{tri:abcsol} split into three inequivalent irreducible
$d$-dimensional representations \eqref{tri:aaasol}, generating nine in total.
\end{Theorem}

In the simplest case, $\lambda^3 = 1$ and $d=k=1$, $\omega=\tau=1$. For three complex units with
$\alpha\beta\gamma=1$, we have $U_\alpha = \alpha$, $U_\beta = \beta$, $U_\gamma = \gamma$ and
equation~\eqref{tri:abcsol} reads as
\[
A = \bmp
0 & 0 & 1 \\
1 & 0 & 0 \\
0 & 1 & 0 \\
\emp,
\qquad
B = \bmp
0 & 0 & \gamma \\
\alpha & 0 & 0 \\
0 & \beta & 0 \\
\emp,
\qquad
C = \lambda\bmp
0 & 0 & \bar{\alpha} \\
\bar{\beta} & 0 & 0 \\
0 & \bar{\gamma} & 0 \\
\emp.
\]

This solution will be reducible if $\alpha=\beta=\gamma=\omega_3^l$.
Since $\mathfrak{Z}_k=1$ (up to $\omega_3$ factor), equations~\eqref{tri:aaasol} give the irreducible parts where $j=0,1,2$,
\[
 A = \omega_3^j,
\qquad B = \omega_3^{j+l},
\qquad C = \lambda \omega_3^{j-l}.
\]

We remark that in general $U_\alpha^d = \alpha^d I$ and similarly for $\beta$, $\gamma$.
This helps to understand how, in a general representation, the tuple $(U,V,W)$ decomposes into
irreducible parts. We will look at the spectral decompositions of $U^d$, $V^d$ and $W^d$.

\section{Canonical order-three unitaries}\label{sec:can}

The matrix $\mathfrak{Z}_k$ is a canonical order-three unitary and
we have just seen its connection to $\Delta{(3,3,3)}$.
In this section, we show that any canonical order-three unitary appears
in representations of the triangle group $\Delta{(3,3,3)}$ and explain
how this can be used.

The following lemma can be checked by a straightforward computation.

\begin{Lemma}\label{can:lem}
 Let $A_M \in {\rm Cli}(d)$ be a canonical order-three unitary with
 $M = \left(\begin{smallmatrix} m_1 & m_2 \\ m_3 & m_4\end{smallmatrix}\right) \in {\rm SL}(2, \mbb{Z}_{\bar{d}})$, so that
 $m_1+m_4 = -1$, $m_1m_4-m_2m_3 = 1$, $M^3 = I$ $\mmod \bar{d}$.
 Then $\forall {\bf a} = (a_1,a_2)\in \mbb{Z}_d^2$
 \begin{align*}
 & (D_{\bf a})\bigl(A_MD_{\bf a}A_M^\dag\bigr)\bigl(A_M^2D_{\bf a}A_M^{2\dag}\bigr)
 = \tau^{-a_1^2m_3+a_1a_2(m_1-m_4)+a_2^2m_2}I, \\
 & (D_{\bf a})\bigl(A_M^2D_{\bf a}A_M^{2\dag}\bigr)\bigl(A_MD_{\bf a}A_M^\dag\bigr)
 = \tau^{-(-a_1^2m_3+a_1a_2(m_1-m_4)+a_2^2m_2)}I.
 \end{align*}
 In particular,
 \begin{align*}
 & (Z)\bigl(A_MZA_M^\dag\bigr)\bigl(A_M^2ZA_M^{2\dag}\bigr) = \tau^{m_2}I, \qquad (Z)\bigl(A_M^2ZA_M^{2\dag}\bigr)\bigl(A_MZA_M^\dag\bigr) = \tau^{-m_2}I.
 \end{align*}
\end{Lemma}

Note that this lemma does not hold for non-canonical order-three unitaries, since
there would be no scalar operator on the right-hand side.
It also can be generalized for $A_{[M, \bf b]}$,
but we will try to keep things simpler here.

Let us put $A = A_M$, $U = \epsilon Z$ and
correspondingly $B = AU$, $C=\lambda U^{-1}A$, where $\epsilon, \lambda \in {\rm S}^1$.

We have that $A^3 = I$ and
\begin{align*}
 &B^3 = \bigl(AUA^{-1}\bigr)\bigl(A^2UA^{-2}\bigr)(U) = \epsilon^3 \tau^{m_2} I , \\
& C^3 = \lambda^3 \bigl(\bigl(A^2UA^{-2}\bigr)\bigl(AUA^{-1}\bigr)(U))^{-1} = \lambda^3 \epsilon^{-3} \tau^{m_2} I, \qquad
 ABC = \lambda I.
\end{align*}

By setting
appropriate $\epsilon$ and $\lambda$, we conclude the following.
\begin{Theorem}\label{can:thm3}
 For a canonical order-three unitary $A_M \in {\rm Cli}(d)$, the tuple
 \[
 (A,B,C) = \bigl(A_M, \epsilon A_MZ, \lambda\epsilon^{-1}Z^{-1}A_M\bigr)
 \]
 with the corresponding
\begin{gather}\label{can:UVW}
 (U,V,W) = (\epsilon Z, \epsilon D_{(-m_2,m_1)}, \epsilon D_{(m_2,m_4)})
\end{gather}
 is a solution of equation~\eqref{tri:ABC} if
 $\epsilon^3 = \tau^{-m_2}$, $\lambda^3 = \omega^{-m_2}$.
\end{Theorem}

One can find that this theorem can be generalized for $A_{[M, {\bf b}]}$ with ${\bf b} \neq 0$,
so that any canonical order-three unitary gives rise to a representation of the triangle group.
But the case of matrices~$A_{[M, {\bf 0}]}$ is principal.
We now use Theorem~\ref{tri:thm} to investigate their properties.

At first, observe that
\begin{gather}\label{can:UW}
 (U,W) = \bigl(\epsilon Z, \epsilon A_M Z A_M^{-1}\bigr)
 = (\epsilon Z, \epsilon D_{(m_2,m_4)})
 = \bigl(\epsilon Z, \epsilon \tau^{m_2m_4} X^{m_2}Z^{m_4}\bigr).
\end{gather}
It follows immediately that if $m_2$ is coprime with $d$
then $U$ and $W$ generate the Weyl--Heisenberg group up to phases,
that is, $\la U,W \ra / \la \xi I \ra = {\rm WH}(d) / \la \xi I \ra$.
Thus $(U,V,W)$ should be irreducible, and since $\mu = \lambda^3 = \omega^{-m_2}$
is a root of~1 of order~$d$, we have the singularity case here.
From equation~\eqref{tri:aaasol}, it follows that there is a~unitary~$T$
that satisfies
\[
 A_M = T\bigl(\omega_3^j \mathfrak{Z}_{-m_2}\bigr)T^\dag,
 \qquad
 \epsilon Z = T (\alpha Z^{-m_2}) T^\dag,
 \qquad
 \epsilon D_{(m_2,m_4)} = T (\alpha D_{(-1,m_2)}) T^\dag
\]
for some $j=0,1,2$ and $\alpha^{3d} = (-1)^{-m_2(d-1)}$.
Hence $T \in {\rm Cli}(d)$ and $A_M$ is conjugate to $\mathfrak{Z}_{-m_2}$
(up to $\omega_3$ factor) in the Clifford group.

If $d > 3$ is prime and $m_2 = 0 \mod d$, then we can consider the conjugation
\[
 \bmp a & 1 \\ 0 & a^{-1} \emp
 \bmp m_1 & 0 \\ m_3 & m_4 \emp
 \bmp a & 1 \\ 0 & a^{-1} \emp^{-1}
 =
 \bmp * & a(m_4-m_1)-m_3 \\ * & * \emp
 \mod d.
\]
We have that $m_4-m_1\neq0 \mmod d$. So there exists $a\in\mbb{Z}_d$~such that $a(m_4-m_1)-m_3=-1 \mmod d$.
And so we have the conjugation to $\mathfrak{Z}_{1}$ by the previous consideration.

Observe that
\[
\bmp l^{-1} & 0 \\ 0 & l \emp
\bmp 0 & -k^{-1} \\ k & -1 \emp
\bmp l & 0 \\ 0 & l^{-1} \emp
=
\bmp 0 & -k^{-1}l^{-2} \\ kl^{2} & -1 \emp
= M_{\mathfrak{Z}_{kl^2}} \mmod d,
\]
hence $\mathfrak{Z}_{k} \simeq \mathfrak{Z}_{kl^2}$, where by ``$\simeq$'' we denote equivalence in the Clifford group.
This means $\mathfrak{Z}_{k} \simeq \mathfrak{Z}_{1}$ if $k\neq0$ is a quadratic residue modulo $d$
and $\mathfrak{Z}_{k} \simeq \mathfrak{Z}_{k_0}$ for some fixed non-residue $k_0$ otherwise.
But we also have
\[
\bmp a & 1 \\ 0 & a^{-1} \emp
\bmp 0 & -1 \\ 1 & -1 \emp
\bmp a^{-1} & -1 \\ 0 & a \emp
=
\bmp * & -\bigl(a^2+a+1\bigr) \\ * & * \emp,
\]
which implies $\mathfrak{Z}_1 \simeq \mathfrak{Z}_{a^2+a+1}$ for any $a\in\mbb{Z}_d$.
If we prove that $a^2+a+1$ can be a quadratic non-residue for some $a$ then
the conclusion is $\mathfrak{Z}_k \simeq \mathfrak{Z}_1$ for any $k\neq0$.
Assume the contrary, that $a^2+a+1$ is a residue for any $a$.
Consider $4\bigl(a^2+a+1\bigr) = (2a+1)^2+3$. This means that $x^2+3$ is a residue for any $x$,
i.e., the set of residues is periodic with period~3, and so any number is a residue,
a contradiction.
We thus showed that for a prime $d > 3$ any canonical order-three unitary $A_M$ is conjugate
to Zauner's unitary $\mathfrak{Z}_1$ in the Clifford group, which was proved in \cite{Fla} by relying
on the number theory more heavily.
If $d=3$, then one can find that $\mathfrak{Z}_1 \not\simeq \mathfrak{Z}_2 \simeq \mathfrak{Z}_1^2$.

In the general case, where $m_2$ and $d$ are not coprime, the described approach still can be useful.
Assume that $n = \gcd(m_2, d)$.
Then \smash{$\mu = \omega^{-m_2} = \omega_{d/n}^{-m_2/n}$}.
Again, let us look at the decomposition of $(U,W)$ from equation~\eqref{can:UW}.
Up to phases, $\la U, W \ra = \la Z, X^{m_2} \ra$ which decomposes into a~sum
of $n$ $d/n$-dimensional irreducible parts.
Let those parts correspond to the parameters
\smash{$\bigl(\alpha_i^{d/n}, \beta_i^{d/n}, \gamma_i^{d/n}\bigr)$},
$i=0,1,\dots,n-1$ as in equation~\eqref{tri:UVWsol}.
By looking at the spectrum of $(\epsilon Z)^{d/n}$, we find that
\[
 \bigcup_{i=0}^{n-1} \bigl\{\epsilon^{d/n} \omega_n^i \bigr\}
 = \bigcup_{i=0}^{n-1} \bigl\{\alpha_i^{d/n}\bigr\}.
\]
Hence, we may think that $\forall i$ $\alpha_i^{d/n} = \epsilon^{d/n} \omega_n^i$.
It follows that those parts are inequivalent.
It is easy to see that they correspond to the invariant subspaces
\ea{\label{can:H}
 H_i = \bigoplus_{l=0}^{d/n-1} \{\ket{i+m_2 l}\}
}
since
\[
 (\epsilon Z)^{d/n}\ket{i+m_2 l}
 = \epsilon^{d/n}\omega_d^{(i+m_2 l)d/n}\ket{i+m_2 l}
 = \epsilon^{d/n}\omega_n^{i}\ket{i+m_2 l}.
\]

Similarly, we have
\begin{align*}
 & (\epsilon D_{(-m_2,m_1)})^{d/n}\ket{i+m_2 l}
 = \epsilon^{d/n}\tau^{-m_2m_1(d/n)^2}\omega_n^{im_1}\ket{i+m_2 l}, \\
 & (\epsilon D_{(m_2,m_4)})^{d/n}\ket{i+m_2 l}
 = \epsilon^{d/n}\tau^{m_2m_4(d/n)^2}\omega_n^{im_4}\ket{i+m_2 l}.
\end{align*}

Recall that $\tau^d = (-1)^{d-1}$. In total, for any $i \in \{0,1,\dots, n-1\}$,
\begin{align*}
 & \bigl(\alpha_i^{d/n}, \beta_i^{d/n}, \gamma_i^{d/n}\bigr) \\
 &\qquad= \bigl(\epsilon^{d/n}\omega_n^{i},
 \epsilon^{d/n}\omega_n^{im_1}(-1)^{(d-1)(m_2/n)m_1(d/n)},
 \epsilon^{d/n}\omega_n^{im_4}(-1)^{(d-1)(m_2/n)m_4(d/n)}\bigr).
\end{align*}
For each $i$, it is either the singularity case, i.e.,
\smash{$\alpha_i^{d/n} = \beta_i^{d/n} = \gamma_i^{d/n}$},
or non-singularity, but those should come in triples which are
shifts of each other.

As an example, let us look at the case where dimension $d=3(3k+1)>3$ is odd and
\[
 M = \bmp 1 & 3 \\ \frac{d-3}{3} & -2 \emp \mod d.
\]
It is known that in these dimensions any canonical order-three unitary is
conjugate to either~$\mathfrak{Z}$,~$\mathfrak{Z}^2$ or $A_M$ \cite{BosWal}.
SIC fiducial vectors were found in the eigenspace of $A_M$ in small dimensions
(this case is denoted by $F_a$ in \cite{SG}).

By Theorem \ref{can:thm3}, we can set $\epsilon = \tau^{-1}$,
$\lambda = \omega^{-1}$ and so $\epsilon^{d/3} = \omega_3$,
\smash{$\mu=\lambda^3 = \omega_{d/3}^{-1}$}.
We have that there are $3$ irreducible $d/3$-dimensional parts
in the decomposition of $(U,V,W)$ that come from equation~\eqref{can:UVW}.
For $i=0,1,2$, we obtain
\[
 \bigl(\alpha_i^{d/3}, \beta_i^{d/3}, \gamma_i^{d/3}\bigr) =
 \bigl(\omega_3^{i+1},
 \omega_3^{i+1},
 \omega_3^{i+1}\bigr),
\]
hence it is the singularity case. This means that invariant subspaces
$H_i$ from equation~\eqref{can:H} are also invariant for $A_M$
and that on each $H_i$ unitary $A_M$ is
conjugate to $\mathfrak{Z}_{-1}$ from ${\rm Cli}(d/3)$.

Another way to see this is to consider the tensor decomposition of $H$ under which
${\rm WH}(d)$ splits as a tensor product of ${\rm WH}(d/3)$ and ${\rm WH}(3)$.
In general, for odd and relatively prime~$n_1$,~$n_2$, we can define
the tensor product $\hat{\otimes}$ for $\mbb{C}^{n_1n_2}$ by
\[
 \ket{i \mmod n_1n_2} = \ket{i \mmod n_1} \hat{\otimes} \ket{i \mmod n_2},
\]
which corresponds to a permutation of the standard basis in $\mbb{C}^{n_1n_2}$.
Under this tensor product, we have that for $a,b \in \mbb{Z}_{n_1n_2}$
\[
 D_{(a,b)}^{(n_1n_2)} = D_{(a,n_2^{-1}b)}^{(n_1)} \hat{\otimes} D_{(a,n_1^{-1}b)}^{(n_2)},
\]
where $n_2^{-1}$ is inverse of $n_2 \mmod n_1$ and $n_1^{-1}$ is inverse of $n_1 \mmod n_2$ (see \cite{Aligned2d, Aligned}).
Conversely, for any $p,q \in \mbb{Z}_{n_1}$, $r,s \in \mbb{Z}_{n_2}$
\[
 D_{(p,q)}^{(n_1)} \hat{\otimes} D_{(r,s)}^{(n_2)} = D_{(pn_2n_2^{-1}+rn_1n_1^{-1},qn_2+sn_1)}^{(n_1n_2)}.
\]
Moreover, under this tensor product the Clifford group splits as well.
One can verify that for~${M' \in {\rm SL}(2, \mbb{Z}_{n_1})}$, $M'' \in {\rm SL}(2, \mbb{Z}_{n_2})$,
we have
\[
A_M = A_{M'} \hat{\otimes} A_{M''}
\]
up to a complex unit factor, where
\ea{\label{can:MMM}
M = n_2 \bmp n_2^{-1} & 0 \\ 0 & 1\emp M' \bmp 1 & 0 \\ 0 & n_2^{-1}\emp
+ n_1 \bmp n_1^{-1} & 0 \\ 0 & 1\emp M'' \bmp 1 & 0 \\ 0 & n_1^{-1}\emp
\in {\rm SL}(2, \mbb{Z}_{n_1n_2}),
}
by looking at the action of $A_M$ on the displacement operators.

In our example, we have $n_1=d/3=3k+1$, $n_2=3$ and thus $n_1^{-1} = 1$, $n_2^{-1} = -k$.
From equation~\eqref{can:MMM}, we conclude that
\[
 M' = \bmp 1 & 0 \\ 0 & -k\emp \left(\bmp 1 & 3 \\ \frac{d-3}{3} & -2 \emp \mmod d/3 \right) \bmp 1 & 0 \\ 0 & 3 \emp
 = \bmp 1 & 9 \\ k & -2\emp \in {\rm SL}(2, \mbb{Z}_{3k+1})
\]
and
\[
 M'' = \bmp 1 & 0 \\ 0 & 1\emp \left(\bmp 1 & 3 \\ \frac{d-3}{3} & -2 \emp \mmod 3 \right) \bmp 1 & 0 \\ 0 & 1 \emp
 = \bmp 1 & 0 \\ 0 & 1 \emp \in {\rm SL}(2, \mbb{Z}_{3}),
\]
implying $A_M = A_{M'} \hat{\otimes} I_3$ for $M' = \left(\begin{smallmatrix} 1 & 9 \\ k & -2\end{smallmatrix}\right) \mmod 3k+1$.
Note that $H_i = \mbb{C}^{d/3} \otimes \ket{i} = \mbb{C}^{d/3} \hat{\otimes} \ket{i}$, hence
$A_M$ acts as the canonical order-three unitary $A_{M'} \in {\rm Cli}(3k+1)$ on each $H_i$
which fits our conclusion obtained before. And we also get that $A_{M'} \simeq \mathfrak{Z}_{-1}$ in an alternative
way.

As a final remark, we add that there is also a notion of
the \textit{full triangle group} $\overline{\Delta}(3,3,3)$.
This group can be obtained from the ordinary $\Delta(3,3,3)$
by adding a reflection along the side of fiducial triangle in Figure~\ref{fig:abc}
(and so any other reflection along triangles sides is there).
In fact, $\big|\overline{\Delta}(3,3,3):{\Delta}(3,3,3)\big| = 2$.
One can see an analogy between reflected motions in
$\overline{\Delta}(3,3,3)$ and anti-unitaries from ${\rm ECli}(d)$.
Perhaps, this connection with motions of the plane could be utilized
for the SIC constructions.

\section{Conclusions}

In this work, we have related canonical order-three unitaries
to representations of the triangle group~${(3,3,3)}$.
We gave the full characterization of those representations
and showed how they can be used to derive properties of
canonical order-three unitaries. In particular, we have proved some
results about them which previously relied on number theory.
We hope this could generate insights about their importance for
the SICs existence problem.

\subsection*{Acknowledgments}
The author is grateful to Ingemar Bengtsson for valuable discussions and
to the referees for important suggestions.
 The author has received funding from the European Union's Horizon 2020 research
 and innovation programme under the Marie Sk{\l}odowska-Curie grant agreement No~873071.
This work was also supported by a grant from the Simons Foundation (1290607,~DY).

\pdfbookmark[1]{References}{ref}
\LastPageEnding

\end{document}